\RequirePackage{fix-cm}
\documentclass[smallextended]{svjour3}       
\smartqed  
\usepackage{graphicx}
\usepackage{amsmath, amssymb, natbib}
\usepackage{xcolor}
\journalname{Celestial Mechanics and Dynamical Astronomy}
\begin{document}

\title{A lower bound of the distance between two elliptic orbits
}
%


\titlerunning{A lower bound of the distance}     

\author{Denis V. Mikryukov         \and
        Roman V. Baluev 
}

\authorrunning{D. V. Mikryukov, R. V. Baluev} 

\institute{D. V. Mikryukov \at
           St. Petersburg State University, Universitetsky pr. 28,
           Stary Peterhof, St. Petersburg 198504, Russia \\
           \email{d.mikryukov@spbu.ru}           
           \and
           R. V. Baluev \at
           St. Petersburg State University, Universitetsky pr. 28,
           Stary Peterhof, St. Petersburg 198504, Russia; \\
           Central Astronomical Observatory at Pulkovo of the
           Russian Academy of Sciences, Pulkovskoje sh. 65/1,
           St. Petersburg 196140, Russia \\
           \email{r.baluev@spbu.ru} 
}

\date{Received: 09 February 2019 / Accepted: 22 May 2019}

\maketitle

\begin{abstract}
We obtain a lower bound of the distance function (MOID) between two noncoplanar bounded
Keplerian orbits (either circular or elliptic) with a common focus. This lower bound is
positive and vanishes if and only if the orbits intersect. It is expressed explicitly,
using only elementary functions of orbital elements, and allows us to significantly
increase the speed of processing for large asteroid catalogs. Benchmarks confirm high
practical benefits of the lower bound constructed.
\keywords{Elliptic orbits \and
          MOID \and
          Linking coefficient \and
          Distance function \and
          Catalogs \and
          Asteroids and comets \and
          Near-Earth asteroids \and
          Space debris \and
          Close encounters \and
          Collisions}
\end{abstract}

\section{Introduction}\label{sec:1}
The problem of computation of a distance between two confocal elliptic
orbits has been intensively studied since the middle of the last century
\citep{Sitarski1968, Vass1978, Dyb1986, Dist, Gron02, Gron05, Armellin2010, Hedo2018}.

In the present article we use the notion \emph{distance} in the sense of the set theory:
minimal value of distances between two points lying on two given confocal ellipses. This
parameter is also known as MOID~--- Minimum Orbital Intersection Distance. From a practical
point of view, the main difficulty of the MOID computation appears due to the lack of the
general analytical solution expressing the result via explicit functions of osculating
elements. A need of numerical methods arises therefore \citep{Gron05, Hedo2018, Baluev2019}.

As a rule, researchers are interested in finding the distance between close orbits. The
precise calculation of the MOID between distant orbits is less relevant. So the problem of
determining a lower bound of the MOID emerged. If the value of this bound proves to be
greater than some positive number $\delta$, then the distance between orbits is greater
than $\delta$ too, and these orbits can be considered safely ``far'' from each other. The
value of closeness threshold $\delta$ depends on the problem considered: which orbital
distance we consider safe, and which is not.

The numeric computation of the MOID, even with fastest algorithms, is relatively expensive
computationally. Therefore the direct comparison between the MOID and the threshold
$\delta$
seems to be impractical, since modern catalogs
typically have a large size. The use of relatively simple lower bound of the distance
between orbits may speed up the selection of hazardously close orbits.

A simple lower bound $\zeta$ of the distance $\rho$ between confocal elliptic orbits
$\mathcal{E}_1$ and $\mathcal{E}_2$ is defined in an inequality
\begin{equation}\label{1-est}
\rho(\mathcal{E}_1, \mathcal{E}_2)\geqslant
\zeta(\mathcal{E}_1, \mathcal{E}_2)\stackrel{\mathrm{def}}{=}\max\{q_1-Q_2, q_2-Q_1\},
\end{equation}
where $q_k$ and $Q_k$ are pericentre and apocentre distances of $\mathcal{E}_k$
respectively. The inequality~\eqref{1-est} holds for any two confocal ellipses
$\mathcal{E}_1$ and $\mathcal{E}_2$, but is informative only if the apocentre of one of the
orbits lies closer to the attracting focus than the pericentre of the other. This is the
case with all eight planets and Pluto except for the pair Neptune~--~Pluto, for which $\zeta<0$.

In practically interesting cases the estimate~\eqref{1-est} usually appears
noninformative, since $\zeta$ becomes negative. A more practical lower bound is presented
in this article. This bound is explicitly expressed through only simple functions of
orbital elements.

The main idea of this lower bound is to construct in the plane of $\mathcal{E}_k$ a
geometrically simple two-dimensional set $\mathcal{H}_k$, containing $\mathcal{E}_k$, and
then to calculate $\rho(\mathcal{H}_1, \mathcal{H}_2)$. The set $\mathcal{H}_k$ is simple
in the sense that it is bounded only by line segments and rays. Enclosing the orbits in
such sets allows one to avoid dealing with the difficult problem of computation of the
distance between second-order curves. The distance between $\mathcal{H}_1$ and
$\mathcal{H}_2$ is easy for an analytic study, since it proves to be equal to the
distance between two skew lines in $\mathbb R^3$, as we will show below. The distance between
these skew lines serves as a positive lower bound of the quantity $\rho(\mathcal{E}_1,
\mathcal{E}_2)$. It never turns negative, and vanishes if and only if $\mathcal{E}_1$ and
$\mathcal{E}_2$ intersect.

\begin{figure}
\includegraphics[width=1\textwidth]{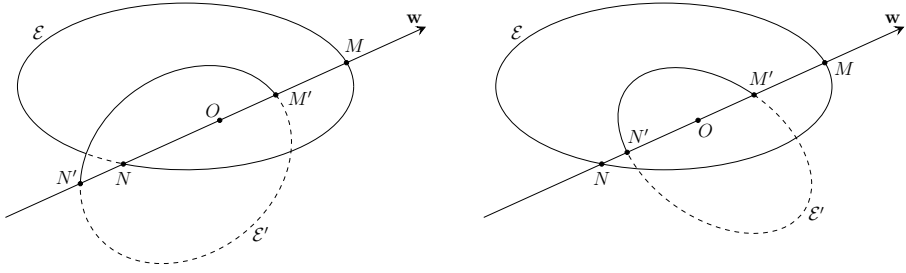}
\caption{The pair of noncoplanar ellipses $\mathcal{E}$ and $\mathcal{E}'$
         can be imbedded in three-dimensional
         space~$\mathbb{R}^3$ in three possible ways. In general, we have
         linked (left) or unlinked (right) configuration. The third degenerate 
         case of intersection separates cases of linked and unlinked~orbits}
\label{fig:1}
\end{figure}

We want to emphasize that in the present article we restrict ourselves to noncoplanar
configurations of elliptic orbits (the notion of skew lines is meaningless in $\mathbb
R^2$). By a pair of elliptic orbits we will always mean two confocal noncoplanar conics,
whose eccentricities belong to a half-open interval $[0; 1)$. We should also notice that
the concept of linked or unlinked orbits \citep{Crowell1963, Link} is essential for our work.
If two orbits $\mathcal{E}_1$ and $\mathcal{E}_2$ have no common points
($\mathcal{E}_1$ and $\mathcal{E}_2$ do not intersect), then they are either linked or unlinked.
Let us recall simple geometric definitions of linked and unlinked configuration of two
noncoplanar elliptic orbits. For this denote by $\mathcal{F}_1$ the plane
containing $\mathcal{E}_1$. The orbits $\mathcal{E}_1$ and $\mathcal{E}_2$ are called
\emph{linked}, if a part of the plane $\mathcal{F}_1$ bounded by the orbit $\mathcal{E}_1$
contains one and only one point belonging to the orbit $\mathcal{E}_2$. If $\mathcal{E}_1$
and $\mathcal{E}_2$ do not satisfy this condition, the orbits $\mathcal{E}_1$ and
$\mathcal{E}_2$ are called \emph{unlinked}. It is easy to see that these definitions are
symmetrical with respect to $\mathcal{E}_1$ and $\mathcal{E}_2$. Continuous transition
between linked and unlinked configurations is possible only through degenerate case of
intersection~(see~Fig.~\ref{fig:1}).

In Section~\ref{sec:2} we formulate the problem in more precise mathematical terms.
In Sections~\ref{sec:3} and~\ref{sec:4} auxiliary geometric constructions are given.
In Section~\ref{sec:5} we obtain the lower bound on the distance, and after that
in Section~\ref{sec:6} we examine its practical efficiency. Section~\ref{sec:7}
provides concluding discussion.

\section{Mathematical setting}\label{sec:2}
Let $\mathcal{E}, \mathcal{E}'\subset\mathbb{R}^3$ be two noncoplanar elliptic orbits with
a common focus $O$, and let Keplerian elements $a, e, i, \omega, \mathrm{\Omega}$ of both orbits refer
to the inertial reference frame $Oxyz$. Elements and all quantities related to~$\mathcal{E}'$
will be marked by a stroke.
Consider the orthogonal unit vectors
\begin{align*}
\mathbf P &= \{ \cos\omega\cos\mathrm{\Omega}-\cos i\sin\omega\sin\mathrm{\Omega},
                \cos\omega\sin\mathrm{\Omega}+\cos i\sin\omega\cos\mathrm{\Omega},
                \sin i\sin\omega\},\\
\mathbf Q &= \{-\sin\omega\cos\mathrm{\Omega}-\cos i\cos\omega\sin\mathrm{\Omega},
               -\sin\omega\sin\mathrm{\Omega}+\cos i\cos\omega\cos\mathrm{\Omega},
                \sin i\cos\omega\}
\end{align*}
and their cross product
$\mathbf Z=\mathbf P\times\mathbf Q=\{\sin i\sin\mathrm{\Omega}, -\sin i\cos\mathrm{\Omega}, \cos i\}$.
Vectors $\mathbf P$, $\mathbf Z$ are parallel to the Laplace--Runge--Lenz vector and
to the angular momentum vector, respectively. For noncoplanar orbits one always has $\sin
I>0$, where $I$ is the angle between $\mathbf Z$ and $\mathbf Z'$. Hence vector $\mathbf
w=\mathbf Z\times\mathbf Z'$ never vanishes for $\mathcal E, \mathcal E'$ and thereby
defines the line of mutual nodes.

The point $O$ decomposes the mutual nodal line into two rays.
A ray whose direction is determined by the vector $\mathbf w$ intersects
$\mathcal{E}$ and $\mathcal{E}'$ at points $M$~and $M'$ respectively. The points
$M$ and $M'$ always exist and are defined uniquely. On the opposite ray one gets
two unique points $N$ and~$N'$~(see~Fig.~\ref{fig:1}). The distance between $\mathcal{E}$
and $\mathcal{E}'$ obviously does not exceed the quantity $\min\{MM', NN'\}$, so
the orbits are made arbitrarily close to each other as either of quantities
$MM'$ or $NN'$ approaches zero. Notice that $MM'$ and $NN'$ can tend to zero
independently of each other and in various ways: one can change the size,
the shape and the spatial orientation of the orbits.

\begin{figure*}
\includegraphics[width=0.75\textwidth]{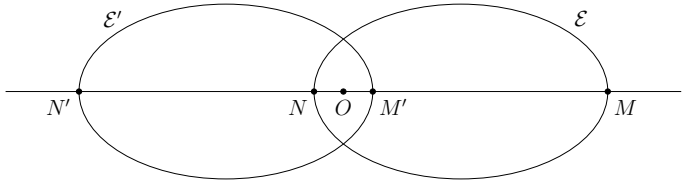}
\caption{Two equal coplanar confocal ellipses with $e=e'=4/5$ and $\mathbf{P}\cdot\mathbf{P}'=-1$}
\label{fig:2}
\end{figure*}

Vanishing of $\min\{MM', NN'\}$ is the necessary and sufficient condition for the
\emph{intersection} of noncoplanar orbits $\mathcal{E}, \mathcal{E}'$, i.e. for
$\mathrm{MOID}=0$. But having $\min\{MM', NN'\}$ small is only a sufficient condition for
the \emph{closeness} of noncoplanar $\mathcal{E}, \mathcal{E}'$ in the MOID sense. In
general, it is not necessary, because MOID can appear small thanks to a small $NM'$ or
$MN'$. Indeed, consider in~Fig.~\ref{fig:2} coplanar orbits $\mathcal{E}$ and $\mathcal{E}'$ with
$a=a'$, $e=e'>1/2$, $\mathbf{P}\cdot\mathbf{P}'=-1$. Let us turn $\mathcal{E}'$ around a common
line of apses through an angle, for example, $\pi/2$. We have
\begin{align*}
      MN' &= 2a(1+e),\\
MM' = NN' &= 2ae,\\
      NM' &= 2a(1-e).
\end{align*}
If $e\to1$ then $MM', NN'\to2a$, $MN'\to4a$, while $NM'\to0$. Since endpoints of $NM'$
lie on different orbits, one concludes that the orbits become arbitrarily close to each other
as $e$ goes to unity. Moving $\mathcal{E}$ and $\mathcal{E}'$ along the common line of apses
towards each other until their distinct foci coincide, we get an analogous example when
making $e\to1$, the quantities $MM'$, $NN'$, $NM'$ tend to positive values, while $MN'\to0$.

We see that any of four quantities
\begin{equation}\label{2-nm}
MM',\;NN',\;MN',\;NM'
\end{equation}
can be made arbitrarily small when the other three remain greater than some predefined
positive value. The line segments $MN$ and $M'N'$ are of no interest here, since they
obviously do not affect the closeness of the orbits.

Put
$$
\sigma_1=r-r',\qquad\sigma_2=R-R',\qquad\sigma_3=r+R',\qquad\sigma_4=R+r',
$$
where $r=OM$, $r'=OM'$ and $R=ON$, $R'=ON'$.
Then
$$
MM'=|\sigma_1|,\qquad NN'=|\sigma_2|,\qquad MN'=\sigma_3,\qquad NM'=\sigma_4.
$$
The quantities $r, r', R, R'$ are easily expressed via osculating elements \citep{Link}
and hence so are~\eqref{2-nm}. While $\sigma_3$ and $\sigma_4$ are always
positive, $\sigma_1$ and $\sigma_2$ can vanish and change the sign. For this reason,
$\sigma_1$ and~$\sigma_2$ carry information about topological configuration of the orbits
$\mathcal{E}$ and~$\mathcal{E}'$. This question is discussed by \citet{Link},
who consider the basic properties of linking coefficient $\ell=\ell(\mathcal{E},
\mathcal{E}')\stackrel{\mathrm{def}}{=}\sigma_1\sigma_2$ of two noncoplanar
orbits\footnote{It is easy to see that $\ell<0$ if and only if $\mathcal{E}$
and $\mathcal{E}'$ are linked, whereas $\ell>0$ if and only if $\mathcal E$,~$\mathcal E'$
are unlinked. The case of zero $\ell$ corresponds to intersection and vice versa.
Thus with the help of the function $\ell$ one can quickly find out which topological configuration
the orbits $\mathcal{E}$ and~$\mathcal{E}'$ have. See \citep{Link} for more details.}.

Let $\mathcal{S}$ and $\mathcal{S}'$ be two arbitrary sets lying in $\mathbb{R}^3$, and let
$Q$ and $Q'$ be two arbitrary points belonging to these sets: $Q\in\mathcal{S}$,
$Q'\in\mathcal{S}'$. By distance between $\mathcal{S}$ and $\mathcal{S}'$ we will always
mean the quantity
\begin{equation}\label{2-rho}
\rho(\mathcal{S}, \mathcal{S}') \stackrel{\mathrm{def}}{=}
\inf_{Q\in\mathcal{S},\,Q'\in\mathcal{S'}}QQ'.
\end{equation}
If $\mathcal{S}$ and $\mathcal{S}'$ are both closed and at least one of them is
bounded (and thus compact), equality~\eqref{2-rho} takes the form
$$
\rho(\mathcal{S}, \mathcal{S}') = \min_{Q\in\mathcal{S},\,Q'\in\mathcal{S'}}QQ'.
$$

Now we can write obvious estimates
\begin{equation}\label{2-estl}
\rho^2(\mathcal{E}, \mathcal{E}')\leqslant|\ell|
\end{equation}
and
\begin{equation}\label{2-ests}
\rho(\mathcal{E}, \mathcal{E}')\leqslant\sigma\leqslant|\ell|^{1/2},
\end{equation}
where
\begin{equation}\label{2-sgm}
\sigma=\sigma(\mathcal{E}, \mathcal{E}')\stackrel{\mathrm{def}}{=}
\min\{|\sigma_1|, |\sigma_2|, |\sigma_3|, |\sigma_4|\}.
\end{equation}
We take the absolute value of $\sigma_3$ and $\sigma_4$ in~\eqref{2-sgm} for the sake of
symmetry. Functions $\ell$ and $\sigma$ are both continuous on the ten-dimensional set of
noncoplanar pairs $(\mathcal{E}, \mathcal{E}')$.

Inequalities~\eqref{2-estl} and~\eqref{2-ests} give simple upper bounds for
$\rho(\mathcal{E}, \mathcal{E}')$. \citet{Link} tried to obtain
a positive lower bound for $\rho(\mathcal{E}, \mathcal{E}')$ with the help of quantities
considered so far. They have shown (see all details in that article) that it is reasonable
to seek this bound in the form of inequality
\begin{equation}\label{2-estcl}
\rho^2(\mathcal{E}, \mathcal{E}')\geqslant C'(e, e', I)|\ell|,
\end{equation}
where $C'$ is a positive function of three real variables $e, e', I$.
Our aim is to solve almost the same problem. We will construct a positive explicit function
$C(e, e', I)$ such that the following inequality is satisfied:
\begin{equation}\label{2-est}
\rho(\mathcal{E}, \mathcal{E}')\geqslant C(e, e', I)\sigma.
\end{equation}

Finding a suitable $C(e, e', I)$ in the estimate~\eqref{2-est} might be important for many
practical applications. Indeed, the right-hand side of~\eqref{2-est} is a simple and
explicit function of osculating elements, so its calculation is much easier than the direct
computation of $\rho(\mathcal{E}, \mathcal{E}')$. The estimate~\eqref{2-est} also allows
one to verify whether two orbits are close to an intersection spending a small CPU time.

\section{Basic geometric constructions}\label{sec:3}
\begin{figure*}
\includegraphics[width=0.75\textwidth]{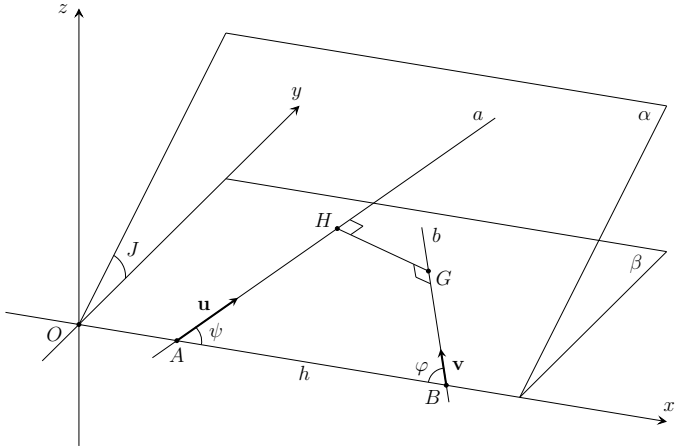}
\caption{The distance $\rho(a, b)$ between the lines $a$ and $b$ depends on $h$ linearly}
\label{fig:3}
\end{figure*}

Let $\alpha, \beta \subset \mathbb{R}^3$ be two two-dimensional closed half-planes that
form a dihedral angle with the plane angle~$J$ satisfying $0<J\leqslant\pi/2$. Introduce a
right-handed Cartesian coordinate system $Oxyz$ in such a way that $\beta=\{y\geqslant0;
z=0\}$ and $\alpha$ lies in a half-space $\{z\geqslant0\}$ (see~Fig.~\ref{fig:3}).
In the positive side of the axis $Ox$ draw points $A$ and $B$ such that $OA<OB$ and
denote $AB=h$. Define lines $a$ and $b$ by vectorial parametric equations
\begin{align*}
\mathbf{r} &= \mathbf{p}+t\mathbf{u},\\
\mathbf{r} &= \mathbf{q}+t\mathbf{v},
\end{align*}
where $t\in\mathbb{R}$, $\mathbf{p}=\overrightarrow{OA}$, $\mathbf{q}=\overrightarrow{OB}$,
\begin{align*}
\mathbf{u} &= \{\cos\psi,\;     \sin\psi\cos J,\; \sin\psi\sin J\},\\
\mathbf{v} &= \{-\cos\varphi,\; \sin\varphi,\;     0\}
\end{align*}
with $0<\varphi, \psi<\pi/2$. We have $A\in a$, $B\in b$, $a\cap\{z\geqslant0\}\subset\alpha$,
$b\cap\{y\geqslant0\}\subset\beta$ (see~Fig.~\ref{fig:3}). The distance between skew lines $a$ and $b$
is given by (see, for example, \cite{Skew})
$$
\rho(a, b)=\frac{|(\mathbf{p}-\mathbf{q})\cdot(\mathbf{u}\times\mathbf{v})|}{|\mathbf{u}\times\mathbf{v}|}.
$$

The distance $\rho(a, b)$ depends only on four arguments $\varphi, \psi, J, h$ and it obviously
tends to zero with $h$. After transformations one obtains
$$
\rho(a, b) = K(\varphi, \psi, J)\,h,
$$
where
\begin{equation}\label{3-k}
\begin{split}
K(\varphi, \psi, J) = \hspace{8cm}\\[8pt]
=\frac{\sin\varphi\sin\psi\sin J}{\sqrt{\sin^2\psi(\sin^2J+\cos^2\varphi\cos^2J)+
                                                            \cos^2\psi\sin^2\varphi+
                                                            \dfrac{\sin2\varphi\sin2\psi\cos J}2}}
\end{split}
\end{equation}
It is easy to check that $K(\varphi, \psi, J)=K(\psi, \varphi, J)$, which stems
from the obvious geometric symmetry. Since $0<J\leqslant\pi/2$, $0<\varphi, \psi<\pi/2$,
the expression under the radical sign in~\eqref{3-k} is always positive.
Let $H\in a$, $G\in b$ be two points such that $\rho(a, b)=HG$. Then $H\in\alpha$
and $G\in\beta$ (see~Fig.~\ref{fig:3}). Furthermore, if any three quantities of
$\varphi, \psi, J, h$ are fixed, then $\rho(a, b)$ tends to zero with the fourth.

\begin{figure*}
\includegraphics[width=0.75\textwidth]{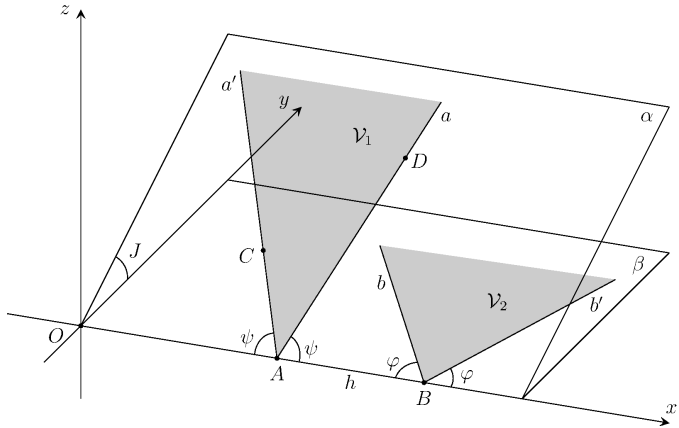}
\caption{The distance $\rho(\mathcal{V}_1, \mathcal{V}_2)$ between
         $\mathcal{V}_1$ and $\mathcal{V}_2$ depends on $h$ linearly,
         since $\rho(\mathcal{V}_1, \mathcal{V}_2)=K(\varphi, \psi, J)\,h$.
         Unlike Figure~\ref{fig:3}, here $a$ and $b$ are (closed) rays}
\label{fig:4}
\end{figure*}

Now, make free our dihedral angle from all the constructions except for the points $A$ and
$B$ lying on its edge. On the face $\alpha$ draw points $C$ and $D$ such that $\angle
CAO=\angle DAB=\psi$, where $0<\psi<\pi/2$. The angle $\angle CAD$ (one assumes $\angle
CAD<\pi$) with its boundary, that is a vertex $A$ and rays $AC$, $AD$, defines in the face
$\alpha$ a two-dimensional closed set $\mathcal{V}_1$ (shaded in~Fig.~\ref{fig:4}). Two-dimensional
closed sets of type $\mathcal{V}_1$ are fundamental to all further constructions. Therefore
for shortness let us call them V-sets. We will define every V-set by its vertex and
exterior angle. For example, we call $\mathcal{V}_1$ a V-set with vertex $A$ and
exterior angle~$\psi$. Any V-set by definition belongs to either of two faces of the
dihedral angle considered. It is always assumed that vertex of any V-set lies on the edge
of the dihedral angle and that exterior angle of any V-set is positive and acute.

Let $\mathcal{V}_2\subset\beta$ be a V-set with vertex $B$ and exterior angle $\varphi$. The
boundary of $\mathcal{V}_2$ is composed from two (closed) rays $b$ and $b'$ emanating from
the vertex~$B$. The boundary of $\mathcal{V}_1$ is also decomposed into two rays $a, a'$
starting from a common origin $A$. We name the rays $a, a', b, b'$ in such a way that
$a'$ and~$b$ intersects the plane $\{x=0\}$ (see~Fig.~\ref{fig:4}).
It is easy to prove (see Appendix) that the distance between $\mathcal{V}_1$ and
$\mathcal{V}_2$ is equal to the distance between straight lines containing the rays $a$
and $b$, so that~$\rho(\mathcal{V}_1, \mathcal{V}_2)=K(\varphi, \psi, J)\,h$.

Suppose that two V-sets having different vertices $Y$ and $Z$ lie in the same
face of the dihedral angle, and have the same exterior angle $\xi$. 
Then, by definition, a two-dimensional union of these V-sets will be called a
W-set with vertices $Y$, $Z$ and exterior angle $\xi$ (the graphical plot is omitted here).

\begin{figure}
\includegraphics[width=0.75\textwidth]{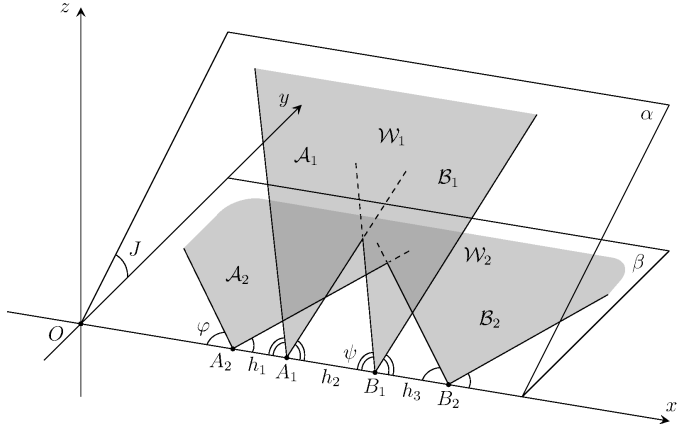}
\caption{Case A). Vertices of a set $\mathcal{W}_1$ both lie
         between vertices of a set $\mathcal{W}_2$.
         The distance $\rho(\mathcal{W}_1, \mathcal{W}_2)$
         tends to zero with $\min\{A_1A_2, B_1B_2, A_1B_2, A_2B_1\}$, which
         remains true if $A_2B_2\subset A_1B_1$}
\label{fig:5}
\end{figure}

\begin{figure}
\includegraphics[width=0.75\textwidth]{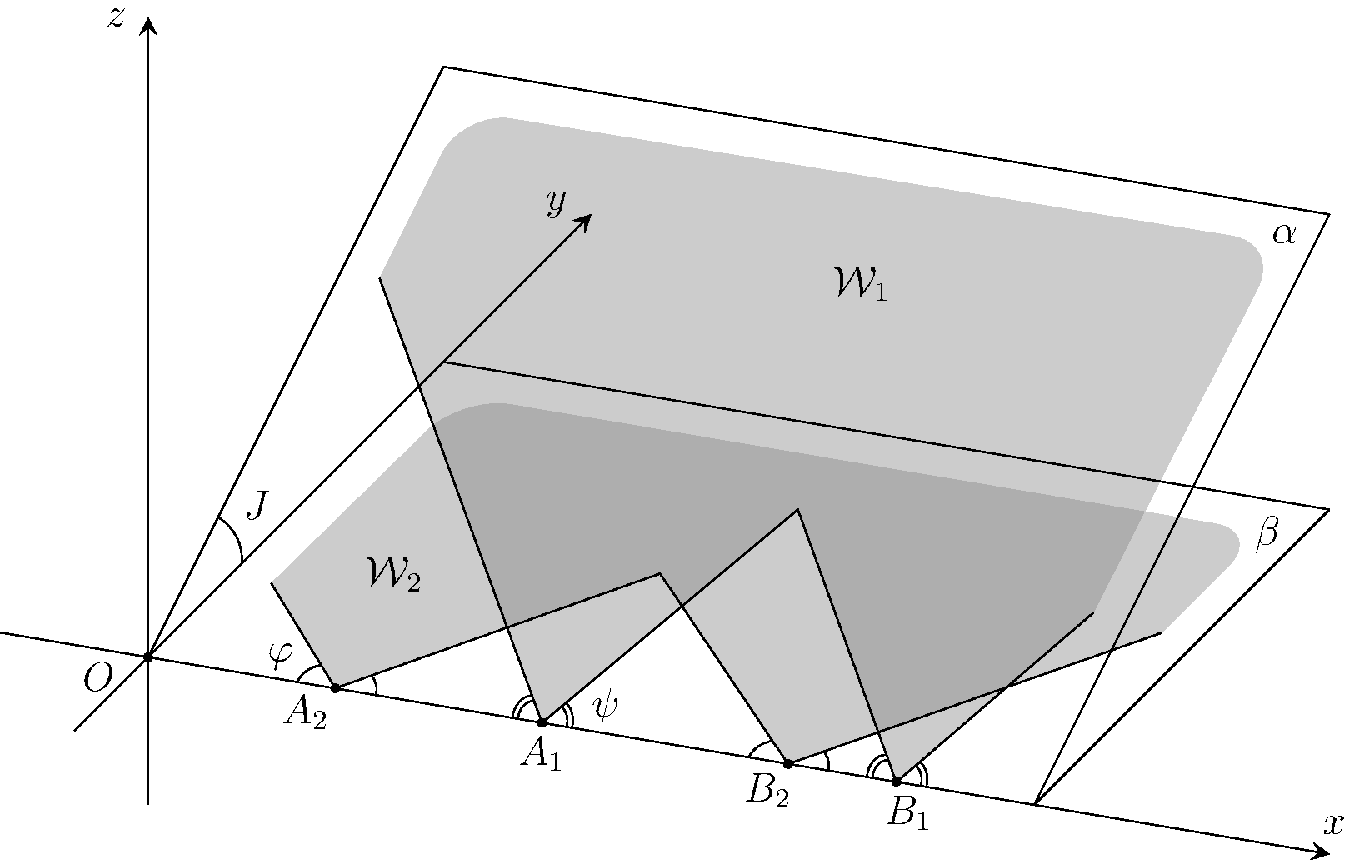}
\caption{Case B). A vertex $A_1$ of a set $\mathcal{W}_1$ lies between
         vertices of a set $\mathcal{W}_2$, but a vertex~$B_1$ does not.
         The distance $\rho(\mathcal{W}_1, \mathcal{W}_2)$ tends to
         zero with $\min\{A_1A_2, B_1B_2, A_1B_2, A_2B_1\}$,
         which remains true if $B_1\in A_2B_2$ and $A_1\notin A_2B_2$}
\label{fig:6}
\end{figure}

Now consider another construction. Draw any four pairwise distinct points $A_1, B_1, A_2,
B_2$ in the positive side of the axis $Ox$. Let $\mathcal{W}_1\subset\alpha$ be a W-set
with vertices $A_1$, $B_1$ and exterior angle $\psi$, and let $\mathcal{W}_2\subset\beta$
be a W-set with vertices $A_2$, $B_2$ and exterior angle~$\varphi$ ($0<\varphi,
\psi<\pi/2$). There are three topologically different possibilities.

A) One of the segments $A_1B_1$ and $A_2B_2$ lies inside another.

B) These segments partly overlap each other (one endpoint of a segment belongs
to another segment, but another endpoint does not).

C) These segments have no common points.

We do not consider case C), since we will not need it anywhere. 

Consider case A). Let first $A_2B_2\supset A_1B_1$, $A_1\in A_2B_1$ (see~Fig.~\ref{fig:5}).
Put $A_2A_1=h_1$, $A_1B_1=h_2$, $B_1B_2=h_3$. Decompose $\mathcal{W}_1$ into two V-sets $\mathcal{A}_1$
and $\mathcal{B}_1$ with vertices $A_1$ and $B_1$ respectively. Analogously, let
$\mathcal{A}_2$ and $\mathcal{B}_2$ be two V-sets with vertices $A_2$ and $B_2$
respectively such that $\mathcal{A}_2\cup\mathcal{B}_2=\mathcal{W}_2$ (see~Fig.~\ref{fig:5}).
The distance between $\mathcal{W}_1$ and $\mathcal{W}_2$ is the smallest of the quantities
\begin{equation}\label{3-rab}
\rho(\mathcal{A}_1, \mathcal{A}_2),\; \rho(\mathcal{A}_1, \mathcal{B}_2),\;
\rho(\mathcal{B}_1, \mathcal{A}_2),\; \rho(\mathcal{B}_1, \mathcal{B}_2).
\end{equation}
Each of four quantities~\eqref{3-rab} is given by $K(\psi, \varphi, J)h$, where $h$
is supposed to be $h_1, h_2+h_3, h_1+h_2, h_3$ respectively. Whence
$$
\rho(\mathcal{W}_1, \mathcal{W}_2)=K(\psi, \varphi, J)\min\{h_1, h_3\}
                                  =K(\psi, \varphi, J)\min\{A_1A_2, B_1B_2\}.
$$
On the other hand, a swap of the points $A_1$ and $B_1$ in~Fig.~\ref{fig:5} gives
$$
\rho(\mathcal{W}_1, \mathcal{W}_2)=K(\psi, \varphi, J)\min\{A_1B_2, A_2B_1\}.
$$
But anyway,
\begin{equation}\label{3-w}
\rho(\mathcal{W}_1, \mathcal{W}_2)=K(\psi, \varphi, J)\min\{A_1A_2, B_1B_2, A_1B_2, A_2B_1\}.
\end{equation}
If $A_1B_1\supset A_2B_2$, the last formula obviously remains true.

Pass to case B) (see Fig.~\ref{fig:6}). Similar to case A) combinatorial considerations
lead to the same formula~\eqref{3-w}.

In view of the above, the general formula for the cases A) and B) is~\eqref{3-w}.
Notice that if any two angles of $\psi, \varphi, J$ are fixed and the third tends to zero,
then in all cases A), B), C) one has $\rho(\mathcal{W}_1, \mathcal{W}_2)\to0$.

\section{Basic constructions on an ellipse}\label{sec:4}
Our goal in this section is to construct a two-dimensional set (see~Sect.~\ref{sec:1})
that necessarily contains the given orbital ellipse. For that, we need to perform
a sequence of geometric constructions layed out below.

On the plane $\mathbb{R}^2$ introduce an inertial right-handed Cartesian coordinate system $Oxy$
and consider on this plane any two different straight lines
$m_1$ and $m_2$. From now on by the angle between the lines $m_1$ and $m_2$ we will
mean the angle between nonoriented lines $m_1$ and $m_2$. Denoting by $\angle(m_1, m_2)$
this angle, one always has $\angle(m_1, m_2)\in[0; \pi/2]$.

\begin{figure}
\includegraphics[width=0.75\textwidth]{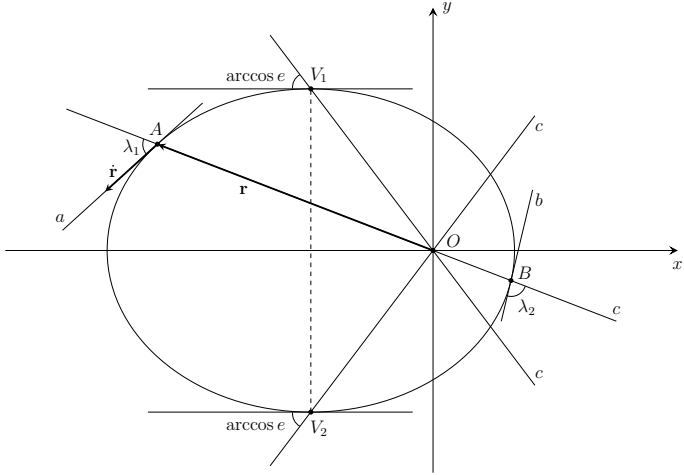}
\caption{The orbital ellipse and auxiliary constructions. The minimum value
         of the angle between the line $c$ and the
         orbit is equal to $\arccos e$, attained only at vertices $V_1$ and
         $V_2$ of semi-minor axes.
         If $0<\theta<\pi$, then $\angle(\mathbf{r},\dot{\mathbf{r}})\in[\arccos e; \pi/2)$.
         If $\pi<\theta<2\pi$, then $\angle(\mathbf{r},\dot{\mathbf{r}})\in(\pi/2; \arccos(-e)]$}
\label{fig:7}
\end{figure}

Let us consider an elliptic orbit with an attracting focus at the origin $O$ and an empty
focus lying in the negative side of the axis $Ox$. Denote by~$e$ the eccentricity
and suppose that the orbit is oriented counterclockwise.
Through the point $O$ draw an arbitrary straight line $c$. We
obtain two points of intersection $A$ and $B$. Draw tangents $a$ and $b$ to the ellipse at
the points $A$ and $B$ respectively (see~Fig.~\ref{fig:7}). In the general case
$\lambda_1\ne\lambda_2$, where $\lambda_1=\angle(a, c)$ and $\lambda_2=\angle(b, c)$. But
if $e\to0$, then $\lambda_1, \lambda_2\to\pi/2$ at any position of the line $c$ (if $e=0$,
one always obviously has $\lambda_1=\lambda_2=\pi/2$). The true anomaly $\theta$ defines
the line $c$ uniquely, and the position of the line $c$ defines the quantity
$\lambda=\min\{\lambda_1, \lambda_2\}$ uniquely. Therefore if we hold $e$ fixed, we may
consider $\lambda$ as a usual function of the true anomaly $\theta$. By
continuity and periodicity, the function $\lambda(\theta)$ necessarily has a maximum and a
minimum. The maximum value is always equal to $\pi/2$ (attained at the apses), while the
minimum depends on $e$. To find the minimum value let us write Cartesian coordinates of the
position and velocity vectors \citep{KholTit}
\begin{align*}
      \mathbf{r} &= \biggl\{\frac{p\cos\theta}{1+e\cos\theta},\quad
                            \frac{p\sin\theta}{1+e\cos\theta}\biggr\},\\
\dot{\mathbf{r}} &= \biggl\{-\sqrt{\frac{\mu}{p}}\,\sin\theta,\quad
                             \sqrt{\frac{\mu}{p}}\,\bigl(e+\cos\theta\bigr)\biggr\},
\end{align*}
where $p$ and $\mu$ are the semi-latus rectum and the gravitational parameter, respectively.
Consider the function of the true anomaly
$$ v(\theta)=\frac{\mathbf{r}\dot{\mathbf{r}}}{|\mathbf{r}||\dot{\mathbf{r}}|}=
             \frac{e\sin\theta}{\sqrt{1+2e\cos\theta+e^2}}
$$
that represents the cosine of the angle between the vectors $\mathbf{r}$ and
$\dot{\mathbf{r}}$. Based on the derivative
$$
v'(\theta)=\frac{e(e+\cos\theta)(1+e\cos\theta)}{(1+2e\cos\theta+e^2)^{3/2}},
$$
we can see that the extrema of $v(\theta)$ are equal to $\pm e$ and are attained at
$\theta=\arccos(-e)$ and $\theta=2\pi-\arccos(-e)$ (at vertices $V_1$ and $V_2$ of
semi-minor axes respectively, see~Fig.~\ref{fig:7}). Hence, the minimum of the periodic function
$\lambda(\theta)$ is equal to $\arccos e$ (see~Fig.~\ref{fig:7}). Now we are able to make the
following remark.

\begin{figure*}
\includegraphics[width=0.75\textwidth]{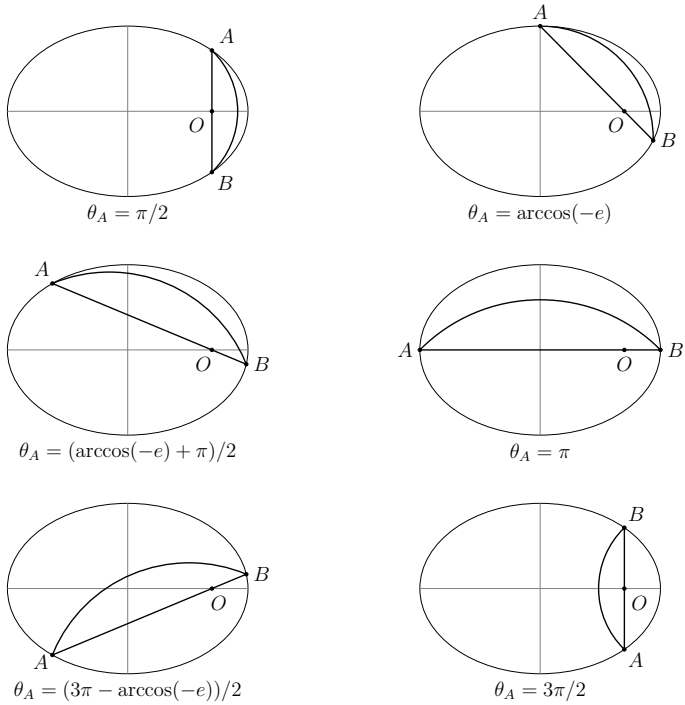}
\caption{Whatever the orientation of the line $c$ is (here are shown some
         particular cases), both arcs of the circumferences lie wholly
         within the ellipse. Lest the figure be overloaded, only one arc
         is shown everywhere. The arc of circumference and the chord $AB$
         are both denoted by the bold line}
\label{fig:8}
\end{figure*}


\begin{remark}\label{rem:1}
Whatever the orientation of the line $c$ is, the angle between the line $c$
and either of two tangents at the points of intersection is not less than $\arccos e$.
\end{remark}

In Remark~\ref{rem:1}, the arbitrariness of the line $c$ (one always
assumes $c\ni O$) is essential, since further this line will play a role of
the mutual line of nodes of two orbits.

Through the points $A$ and $B$ let us draw a circumference in such a way that $\angle(c, m)
= \angle(c, n) = \arccos e$, where $m$ and $n$ are tangents to the circumference at the
points $A$ and $B$ respectively. Such construction can be done in two possible ways, so
that we obtain two (equal) circumferences sharing two common points~$A$ and~$B$. We need
only the shorter arcs of these circumferences subtended by a chord~$AB$~(see~Fig.~\ref{fig:8}).

\begin{remark}\label{rem:2}
Whatever the orientation of the chord $AB$ is, all interior points of both arcs lie inside
the ellipse.
\end{remark}

Remark~\ref{rem:2} follows from Remark~\ref{rem:1}. The validity of
Remark~\ref{rem:2} can also be ascertained by usual means of analytic geometry.

\begin{figure}
\includegraphics[width=0.75\textwidth]{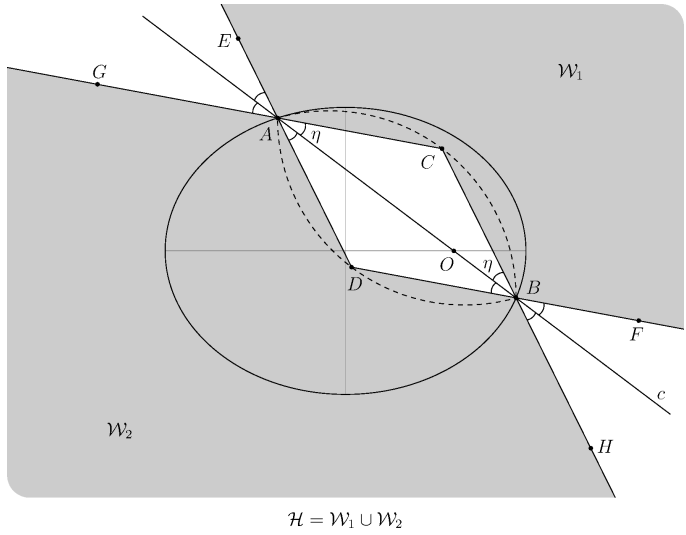}
\caption{Polygonal chains $EACBF$ and $GADBH$ define in the orbit plane
         W-sets $\mathcal{W}_1$ and $\mathcal{W}_2$ respectively.
         A two-dimensional union $\mathcal{H}$ of these
         contains the whole orbit,
         whatever the relative position of
         the line $c$ and the orbit is}
\label{fig:9}
\end{figure}

On the segment $AB$ as a base draw an isosceles triangle~$\bigtriangleup ACB$, whose
apex $C$ lies on either of two arcs considered (see~Fig.~\ref{fig:9}, where both of these arcs are
dashed). A base angle $\eta$ is easily calculated and equals to $\eta=(\arccos e)/2$. Draw two
rays $AE$ and $BF$ such that $AE\parallel BC$, $BF\parallel AC$, and that points $C, E, F$
lie on one side of the line $c$~(see~Fig.~\ref{fig:9}). With the help of
Remark~\ref{rem:1}, one establishes that all interior points of the rays $AE$ and
$BF$ lie \emph{outside} of the ellipse, while Remark~\ref{rem:2} guarantees that
all interior points of the line segments $AC$ and $CB$ lie \emph{inside} the ellipse.
A polygonal chain $EACBF$ on the plane $Oxy$ defines a W-set $\mathcal{W}_1$ with vertices
$A, B$ and exterior angle $\eta$. On the other side of the line~$c$ we construct in an
analogous way a W-set $\mathcal{W}_2$ with vertices $A, B$ and exterior angle $\eta$,
that is defined in Fig.~\ref{fig:9} by a polygonal chain $GADBH$.

Given any orientation of the line $c$, the orbit is completely contained in a set
$\mathcal{H}=\mathcal{W}_1\cup\mathcal{W}_2$. The size of the ellipse, its shape and the
position of the line $c$ define the set $\mathcal{H}$ uniquely. Further two-dimensional
closed sets of type $\mathcal{H}$ will be called by H-sets. Every H-set will be defined by
vertices and exterior angle of those two (equal) W-sets, that give this H-set. For
example, $\mathcal{H}$ is an H-set with vertices $A, B$ and exterior angle $\eta$.

Consider some properties of H-sets defined above.
Every H-set is pathwise connected. The same can be said about its boundary.
By rotating the line~$c$ about the focus $O$, we obtain different
H-sets. But if $e$ is fixed, then all of these H-sets are similar: the only difference is
the distance between the vertices. Exterior angle~$\eta$ of any H-set satisfies
$0<\eta\leqslant\pi/4$. The maximum value $\pi/4$ is attained only for circular orbits. In
this case a rhombus $ACBD$ (see~Fig.~\ref{fig:9}) turns into a square inscribed in a circular orbit,
while this orbit itself can be considered as the result of degeneration of the
circumference arcs (dashed in~Fig.~\ref{fig:9}) into a union of two semicircumferences with common
extremities $A$ and $B$.

\section{The lower bound of the distance between orbits}\label{sec:5}
Return to the notations of the Section~\ref{sec:2} and again consider two noncoplanar elliptic
orbits~$\mathcal{E}$ and~$\mathcal{E}'$ with a common focus $O$ (see~Fig.~\ref{fig:1}).
Denote by $c$ the mutual nodal line.

In the plane of the orbit $\mathcal{E}$
define an H-set $\mathcal{H}$ with vertices $M, N$ and exterior angle
\begin{equation}\label{5-et1}
\eta=\frac{\arccos e}2.
\end{equation}
In the plane of the orbit $\mathcal{E}'$ define an H-set $\mathcal{H}'$ with
vertices $M', N'$ and exterior angle
\begin{equation}\label{5-et2}
\eta'=\frac{\arccos e'}2.
\end{equation}
Since $\mathcal{E}\subset\mathcal{H}$, $\mathcal{E}'\subset\mathcal{H}'$, one has
\begin{equation}\label{5-eh}
\rho(\mathcal{E}, \mathcal{E}')\geqslant\rho(\mathcal{H}, \mathcal{H}').
\end{equation}
The planes of the orbits $\mathcal{E}$ and $\mathcal{E}'$ decompose all the space
$\mathbb{R}^3$ into four dihedral angles. Divide~$\mathcal{H}$ and~$\mathcal{H}'$ into
W-sets $\mathcal{W}$, $\mathcal{W}'$, $\mathcal{Y}$, $\mathcal{Y}'$ such that
$\mathcal{H}=\mathcal{W}\cup\mathcal{Y}$, $\mathcal{H}'=\mathcal{W}'\cup\mathcal{Y}'$. We
name these four W-sets in such a way that $\mathcal{W}$ and $\mathcal{W}'$ lie in the faces
of that dihedral angle, whose plane angle does not exceed~$\pi/2$ (so that
$\mathcal{Y}$ and $\mathcal{Y}'$ fall into the faces of another dihedral angle, whose
plane angle does not exceed~$\pi/2$). The sets $\mathcal{H}$ and $\mathcal{H}'$ both have
axial symmetry around the axis $c$, which implies $\rho(\mathcal{H},
\mathcal{H}')=\rho(\mathcal{W}, \mathcal{W}')$. Whence by~\eqref{5-eh} we obtain an
estimate
\begin{equation}\label{5-rw}
\rho(\mathcal{E}, \mathcal{E}')\geqslant\rho(\mathcal{W}, \mathcal{W}').
\end{equation}

If the orbits $\mathcal{E}$ and $\mathcal{E}'$ do not intersect, then they are either
linked or unlinked~(see~Fig.~\ref{fig:1}). Suppose first that $\mathcal{E}$ and $\mathcal{E}'$
are unlinked. Then one of the line segments $MN$ and $M'N'$ is completely contained in another one
(see~Fig.~\ref{fig:1}, right). Thus the relative position of $\mathcal{W}$ and $\mathcal{W}'$
for unlinked orbits corresponds to the case A) of Section~\ref{sec:3}.
Further, if $\mathcal{E}$ and $\mathcal{E}'$ are linked, the line segments $MN$ and $M'N'$
partly overlap each other (one endpoint of the segment $MN$ belongs to $M'N'$, but another
one does not, see~Fig.~\ref{fig:1}, left). So for linked orbits the relative position of
$\mathcal{W}$ and $\mathcal{W}'$ corresponds to the case~B) of Section~\ref{sec:3}.
Using the general formula~\eqref{3-w} for cases A) and B) and taking into account that
unlike the angle $J$ (see Section~\ref{sec:3}), the angle $I$ is allowed
to lie in the second quadrant\footnote{Note that the equality
$\cos I=\mathbf{Z}\cdot\mathbf{Z}'=\cos i\cos i'+\sin i\sin i'\cos(\mathrm{\Omega}-\mathrm{\Omega}')$ defines
the angle~$I$ uniquely, since $0<I<\pi$.}, we obtain
\begin{equation}\label{5-rks}
\rho(\mathcal{W}, \mathcal{W}')=K(\eta, \eta', \min\{I, \pi-I\})\sigma,
\end{equation}
where $K, \eta, \eta', \sigma$ are defined by~\eqref{3-k},~\eqref{5-et1},
\eqref{5-et2},~\eqref{2-sgm} respectively. According to~\eqref{5-rw} and~\eqref{5-rks}
one obtains the final inequality
\begin{equation}\label{5-finest}
\rho(\mathcal{E}, \mathcal{E}')\geqslant C(e, e', I)\sigma,
\end{equation}
where
$$
C(e, e', I)=K\Bigl(\frac{\arccos e}2, \frac{\arccos e'}2, \min\{I, \pi-I\}\Bigr).
$$
After transformations we obtain
\begin{equation}\label{5-c}
C(e, e', I) = \sqrt{  \frac  {(1-e)(1-e')\sin^2\!I}
                             {(1-e)(1-e')\sin^2\!I+2(1+|\cos I|\sqrt{(1-e^2)(1-e'^2)}-ee')}
                   }.
\end{equation}
For any noncoplanar $\mathcal{E}$ and $\mathcal{E}'$, that is when $\sin I>0$,
the function~\eqref{5-c} is positive and satisfies $C(e, e', I)=C(e', e, I)$.

Inequalities~\eqref{2-ests} and~\eqref{5-finest} give an effective bilateral estimate
\begin{equation}\label{5-est}
\tau(\mathcal{E}, \mathcal{E}')\leqslant\rho(\mathcal{E}, \mathcal{E}')
                               \leqslant\sigma(\mathcal{E}, \mathcal{E}'),
\end{equation}
where we have put by definition
\begin{equation}\label{5-tau}
\tau(\mathcal{E}, \mathcal{E}')=C(e, e', I)\sigma(\mathcal{E}, \mathcal{E}').
\end{equation}
If $\sin I>0$, then the functions $\tau, \rho, \sigma$ are either
together equal to zero ($\mathcal{E}, \mathcal{E}'$ intersect) or
together positive ($\mathcal{E}, \mathcal{E}'$ do not intersect).
The estimate~\eqref{5-est} of the distance $\rho(\mathcal{E}, \mathcal{E}')$
contains only simple and explicit functions of osculating elements.
Note that the lower estimate~\eqref{5-finest},~\eqref{5-c} formally remains true for coplanar orbits too.
Indeed, from continuity of the function $C(e, e', I)$ and boundedness of the function
$\sigma(\mathcal{E}, \mathcal{E}')$ (see definitions~\eqref{5-c} and~\eqref{2-sgm}
respectively) it follows that whenever $\sin I=0$ the estimate~\eqref{5-finest} turns
into a noninformative but always valid inequality $\rho(\mathcal{E}, \mathcal{E}')\geqslant0$.

\section{On the practical efficiency of the estimate constructed}\label{sec:6}
We used an asteroid orbits database of the Minor Planet Center (MPC), downloaded from the
official site \texttt{https://minorplanetcenter.net} on November~10,~2018. On that date
this database contained~$523\,824$ numbered objects. We used the first $20\,000$ of these 
for constructing three different catalogs $\mathrm{\Phi}_1, \mathrm{\Phi}_2, \mathrm{\Phi}_3$ of orbit pairs. Each
catalog has been constructed in accordance with the given value~$\delta$ of an upper
threshold of the distance between two orbits $\mathcal{E}$ and~$\mathcal{E}'$
(see~Table~\ref{tab:1}). Namely, of all~$N_0=199\,990\,000$ pairs
in the original sample $\mathrm{\Phi}$, we put in each
catalog those and only those pairs~$\mathcal{E}$ and~$\mathcal{E}'$ that satisfied the
condition $\rho(\mathcal{E}, \mathcal{E}')\leqslant\delta$. The computation of the distance
$\rho(\mathcal{E}, \mathcal{E}')$ was carried out by means of the software described by
\citet{Baluev2019} and available for download at
\texttt{http://sourceforge.net/projects/distlink}. This software provides a numeric
implementation of the algebraic method presented by \citet{Dist},
similar to the one presented by \citet{Gron02, Gron05}. All calculations in our work have
been carried out with an Intel Core i5-4460 PC @ 3.2GHz with 7.7GiB of RAM.

\begin{table}
\caption{The characteristics of the catalogs $\mathrm{\Phi}_1, \mathrm{\Phi}_2, \mathrm{\Phi}_3$.
         The number $N_0=199\,990\,000$. See text for details}
\label{tab:1}
\begin{tabular}{lllllllll}
\hline\noalign{\smallskip}
$\mathrm{\Phi}_k$ & $\delta$, AU & $N$ & $N_{\mathrm{skip}}$ & $N_0-N_{\mathrm{skip}}$
                          & $\frac{N_{\mathrm{skip}}}{N_0}\cdot100\%$
                          & $T_C$, s & $T$, s & $T/T_C$\\
\noalign{\smallskip}\hline\noalign{\smallskip}
$\mathrm{\Phi}_1$  & $0.01$ & $8\,137\,922$ & $127\,546\,890$ & $72\,443\,110$ & $63.77\%$
                                   & $2\,770$ & $6\,685$ & $2.41$ \\
$\mathrm{\Phi}_2$  & $0.005$ & $4\,147\,316$ & $157\,245\,149$ & $42\,744\,851$ & $78.62\%$
                                   & $1\,730$ & $6\,554$ & $3.78$ \\
$\mathrm{\Phi}_3$  & $0.0026$ & $2\,180\,010$ & $175\,301\,105$ & $24\,688\,895$ & $87.65\%$
                                   & $1\,100$ & $6\,515$ & $5.92$ \\
\noalign{\smallskip}\hline
\end{tabular}
\end{table}

For catalogs $\mathrm{\Phi}_1, \mathrm{\Phi}_2, \mathrm{\Phi}_3$ the values of $\delta$ were chosen to be
approximately four~($0.01$~AU), two~($0.005$~AU) and one~($0.0026$~AU) Earth--Moon
distances, respectively (see~Table~\ref{tab:1}). Clearly, the smaller the value of~$\delta$, the
smaller the number $N$ of asteroid pairs contained in $\mathrm{\Phi}_k$ catalog. Thus we have
$\mathrm{\Phi}\supset\mathrm{\Phi}_1\supset\mathrm{\Phi}_2\supset\mathrm{\Phi}_3$. Each catalog
has been constructed two times in different ways.

The first way to build $\mathrm{\Phi}_1, \mathrm{\Phi}_2, \mathrm{\Phi}_3$ was to calculate
$\rho(\mathcal{E}, \mathcal{E}')$ for \emph{each} orbit pair in $\mathrm{\Phi}$.
If a pair from~$\mathrm{\Phi}$ satisfied $\rho(\mathcal{E}, \mathcal{E}')\leqslant\delta$,
then it was put in the catalog; otherwise, it was skipped. According to~Table~\ref{tab:1},
the average computation time $T$ of building each catalog in such a way
is approximately $6\,600$ seconds. Whence the average computation time $t_{\mathrm{MOID}}$
of $\rho(\mathcal{E}, \mathcal{E}')$ per one pair is
\begin{equation}\label{6-tm}
t_{\mathrm{MOID}}\approx\frac{6\,600\cdot10^6}{N_0}\;
\mbox{$\mathrm{\mu s}$}\approx33\;\mbox{$\mathrm{\mu s}$}.
\end{equation}

After that, the same catalogs $\mathrm{\Phi}_1, \mathrm{\Phi}_2, \mathrm{\Phi}_3$ were built in the second way,
which uses the estimate~\eqref{5-finest},~\eqref{5-c}. For each pair from $\mathrm{\Phi}$
the function $\tau(\mathcal{E}, \mathcal{E}')$ was initially calculated. The distance
$\rho(\mathcal{E}, \mathcal{E}')$ was calculated if and only if
$\tau(\mathcal{E}, \mathcal{E}')$ satisfied
\begin{equation}\label{6-csd}
\tau(\mathcal{E}, \mathcal{E}')\leqslant\delta.
\end{equation}
The pair of $\mathcal{E}$, $\mathcal{E}'$ was written in the catalog if and
only if it satisfied $\rho(\mathcal{E}, \mathcal{E}')\leqslant\delta$. The
time~$T_C$ of constructing the catalog $\mathrm{\Phi}_k$ in the second way is expected to be less than
the time $T$ of constructing the same $\mathrm{\Phi}_k$ in the first way, since in the
second case $\rho(\mathcal{E}, \mathcal{E}')$ is computed \emph{not for all} pairs
from $\mathrm{\Phi}$ (but only for those satisfying~\eqref{6-csd}). Obviously, the less $\delta$,
the more significant difference between $T$ and $T_C$ should be.
Table~\ref{tab:1}, where we present our values of $T_C$ for each $\mathrm{\Phi}_k$,
confirms these evident assumptions.

\begin{table}
\caption{The values of $\tau, \rho, \sigma$ for some close orbits
         in the Main Belt. Six true digits after the decimal point
         are indicated in each decimal fraction}
\label{tab:2}
\begin{tabular}{llll}
\hline\noalign{\smallskip}
Pair of orbits $\mathcal{E}$ and $\mathcal{E}'$ & $\tau(\mathcal{E},  \mathcal{E}')$, AU
                                                & $\rho(\mathcal{E},  \mathcal{E}')$, AU
                                                & $\sigma(\mathcal{E},\mathcal{E}')$, AU\\
\noalign{\smallskip}\hline\noalign{\smallskip}
14 Irene\;       --\; 32 Pomona       & $0.001121$  & $0.009954$ & $0.010824$ \\
4 Vesta\;        --\; 17 Thetis       & $0.000104$  & $0.004216$ & $0.004709$ \\
722 Frieda\;     --\; 1218 Aster      & $0.000123$  & $0.001106$ & $0.005698$ \\
946 Poesia\;     --\; 954 Li          & $0.000132$  & $0.000968$ & $0.009371$ \\
704 Interamnia\; --\; 775 Lumiere     & $0.000068$  & $0.000696$ & $0.001000$ \\
1 Ceres\;        --\; 512 Taurinensis & $0.000017$  & $0.000180$ & $0.000504$ \\
1333 Cevenola\;  --\; 4699 Sootan     & $0.000006$  & $0.000033$ & $0.000036$ \\
\noalign{\smallskip}\hline
\end{tabular}
\end{table}

In Table~\ref{tab:2}, we give for some orbit pairs from $\mathrm{\Phi}_1$ our values
of the functions that are in the estimate~\eqref{5-est}.

Let us try to approximately determine how many times the average time $t_{\mathrm{EST}}$
spent on the calculating the function $\tau(\mathcal{E}, \mathcal{E}')$ and verifying
the condition~\eqref{6-csd} (per one pair) less than the average time $t_{\mathrm{MOID}}$ computed above.
To do so, notice that when $\mathrm{\Phi}_k$ is calculated in the second way the computation of
$\tau(\mathcal{E}, \mathcal{E}')$ and verifying~\eqref{6-csd} are performed for all $N_0$
pairs from~$\mathrm{\Phi}$. But the distance $\rho(\mathcal{E}, \mathcal{E}')$ is computed
only for $N_0-N_{\mathrm{skip}}$ orbit pairs, where~$N_{\mathrm{skip}}$ is a number
of pairs that have not satisfied~\eqref{6-csd} (a number of skipped pairs, where orbits
are definitely distant from each other). So the time $T_C$ is roughly made up
of two parts: $N_0\cdot t_{\mathrm{EST}}$ (calculating
$\tau(\mathcal{E}, \mathcal{E}')$ and verifying~\eqref{6-csd} for every pair from~$\mathrm{\Phi}$) and
$(N_0-N_{\mathrm{skip}})\cdot t_{\mathrm{MOID}}$ (computation of $\rho(\mathcal{E}, \mathcal{E}')$
only for potentially close orbits)\footnote{A more precise calculation should take into account, for example,
accompanying file IO operations.}. We obtain an equation
\begin{equation}\label{6-nt}
N_0\cdot t_{\mathrm{EST}}+(N_0-N_{\mathrm{skip}})\cdot t_{\mathrm{MOID}}=T_C,
\end{equation}
where $N_{\mathrm{skip}}$ and $T_C$ are supposed to correspond to the same
$\mathrm{\Phi}_k$~(see~Table~\ref{tab:1}). From~\eqref{6-nt},~\eqref{6-tm} one obtains
$t_{\mathrm{EST}}\approx1.9$ $\mathrm{\mu s}$, $t_{\mathrm{EST}}\approx1.6$ $\mathrm{\mu s}$,
$t_{\mathrm{EST}}\approx1.4$ $\mathrm{\mu s}$ for $\mathrm{\Phi}_1$, $\mathrm{\Phi}_2$,
$\mathrm{\Phi}_3$ respectively, whence by~\eqref{6-tm} finally
$$
17\lesssim\frac{t_{\mathrm{MOID}}}{t_{\mathrm{EST}}}\lesssim23.
$$
We see that a selection criterion of potentially close orbits based on the
computation of the function $\tau(\mathcal{E}, \mathcal{E}')$ and comparing it
with some threshold value~$\delta$ is processed approximately twenty times
faster than the direct computation of~$\rho(\mathcal{E}, \mathcal{E}')$.
These figures are rather rough, but even so they clearly show what computational
benefits can be gained when using the estimate~\eqref{5-finest},~\eqref{5-c}.

The values of timedimensional quantities $t_\mathrm{MOID}$ and $t_\mathrm{EST}$
are heavily dependent on the hardware used.
The same can be said about $T$ and $T_C$~(see~Table~\ref{tab:1}).
However if $\rho(\mathcal{E}, \mathcal{E}')$ is computed by the same software,
the dimensionless quantity $t_\mathrm{MOID}/t_\mathrm{EST}$
should keep approximately the same value. When other software is used
\citep[see for example,][]{Gron05, Hedo2018}, the value of~$t_\mathrm{MOID}/t_\mathrm{EST}$
may differ significantly from our one. Indeed, benchmarking tests carried out in our
previous article \citep{Baluev2019} reveal that computational performance (time of
calculating $\rho(\mathcal{E}, \mathcal{E}')$ per one pair) vary considerably from one
software to another \citep[see also][]{Hedo2018}. On the other hand, the computation
of the estimate~\eqref{5-finest},~\eqref{5-c} 
is not a numeric issue for any computation MOID library. We conclude that
the slower the MOID computation numeric algorithm (when testing on the same hardware and
with the same precision), the larger the quantity $t_\mathrm{MOID}/t_\mathrm{EST}$ should
be. In contrast, the time ratio~$T/T_C$ is expected to be less
susceptible to a change of the software used, 
because $T/T_C$ mainly depends only on the number of skipped pairs $N_\mathrm{skip}$.
Again, this our conclusions regarding the quantities~$t_\mathrm{MOID}/t_\mathrm{EST}$,
$T/T_C$ and their dependence on the concrete MOID numeric library are rather empirical
and need more thorough and complete investigation.

\section{Discussion}\label{sec:7}
With the result presented above we are able to quickly compute the two-sided range 
for the MOID without computing the MOID itself. The lower bound of the MOID is rather novel result,
and it is probably more important for practical applications than the upper one.
However, the efficiency of this bound still needs to be discussed.

\begin{table}
\caption{Four pairs from the original sample $\mathrm{\Phi}$ with the largest value of $\tau/\rho$}
\label{tab:max}
\begin{tabular}{lll}
\hline\noalign{\smallskip}
Pair of orbits from $\mathrm{\Phi}$  & $\tau/\rho$ & $I$             \\
\noalign{\smallskip}\hline\noalign{\smallskip}
3873 Roddy\;    --\; 5496 1973 NA    & $0.529708$  & $88.3709^\circ$ \\
5496 1973 NA\;  --\; 17408 McAdams   & $0.529070$  & $83.2490^\circ$ \\
2063 Bacchus\;  --\; 3200 Phaethon   & $0.524568$  & $28.9738^\circ$ \\
5496 1973 NA\;  --\; 5869 Tanith     & $0.523234$  & $79.3475^\circ$ \\
\noalign{\smallskip}\hline
\end{tabular}
\end{table}

\begin{table}
\caption{The elements of the orbits 5335 Damocles and 31824 Elatus on November~10,~2018
         according to MPC database}
\label{tab:elem}
\begin{tabular}{llllll}
\hline\noalign{\smallskip}
Orbit & $a$, AU & $e$ & $i$ & $\omega$ & $\mathrm{\Omega}$ \\
\noalign{\smallskip}\hline\noalign{\smallskip}
5335 Damocles   & $11.8305615$ & $0.8663989$ & $61.68564^\circ$ & $191.27338^\circ$ & $314.05405^\circ$\\
31824 Elatus    & $11.7977758$ & $0.3813889$ & $5.24419^\circ$  & $281.43833^\circ$ & $87.18966^\circ$\\
\noalign{\smallskip}\hline
\end{tabular}
\end{table}

\begin{table}
\caption{The values of $\tau$, $\rho$, $\sigma$ and $I$ for configuration 5335 Damocles -- 31824 Elatus.
         The input elements are in Table~\ref{tab:elem}}
\label{tab:cent}
\begin{tabular}{llllll}
\hline\noalign{\smallskip}
$\tau$, AU & $\rho$, AU & $\sigma$, AU & $\tau/\rho$ & $\sigma/\rho$ & $I$ \\
\noalign{\smallskip}\hline\noalign{\smallskip}
$1.862996$  & $3.200890$ & $9.547594$ & $0.582024$ & $2.982793$ & $65.3353^\circ$\\
\noalign{\smallskip}\hline
\end{tabular}
\end{table}

The lower bound $\tau$ for $\rho$ is hardly optimal. Consider for example two
circular perpendicular orbits ($e=e'=0$, $I=\pi/2$). The
relations~\eqref{5-finest},~\eqref{5-c}~give
$$
\rho\geqslant\tau=\frac{\sigma}{\sqrt{3}}\approx0.577\sigma,
$$
though it is clear that in this case $\rho=\sigma$.
The values of~$\tau$,~$\rho$ given in~Table~\ref{tab:2} also suggest that the lower bound
$\tau$ is not optimal. We calculated the ratio $\tau/\rho$ for all $N_0=199\,990\,000$
pairs in the original sample $\mathrm{\Phi}$ and collected in Table~\ref{tab:max} four
pairs having the largest (among all $N_0$ pairs) values of $\tau/\rho$. As~Table~\ref{tab:max}
indicates, none of the pairs from $\mathrm{\Phi}$ show an inequality $\tau/\rho\geqslant0.53$.
This result is even worse than $\tau/\rho=1/\sqrt3\approx0.577$ that corresponds to circular
perpendicular case considered above. Nevertheless, during our experiments with the whole database
MPC we managed to find one pair of real orbits for which $\tau/\rho>1/\sqrt3$. Centaurs Damocles
and Elatus (catalog numbers 5335 and 31824 respectively) get $\tau/\rho\approx0.582$.
In~Table~\ref{tab:elem} we give the elements of these orbits, and in~Table~\ref{tab:cent}
we present the main characteristics of their configuration. Is there an example
of a pair of orbits (real or simulated) with $\tau/\rho\geqslant3/5$? So far we
have never seen such configurations, but our opinion is that these must exist. Perhaps
further large scale experiments will reveal\footnote{Notice that cometary mutual elements
proposed by \citet{Gron05} may prove to be more suitable for large scale experiments than usual orbital
elements given in Table~\ref{tab:elem}.} orbital configurations showing $\tau/\rho\geqslant3/5$.

\begin{table}
\caption{The arithmetic mean of the quantity $\tau/\rho$ for catalogs $\mathrm{\Phi}$,
         $\mathrm{\Phi}_1$, $\mathrm{\Phi}_2$, $\mathrm{\Phi}_3$}
\label{tab:mean}
\begin{tabular}{lll}
\hline\noalign{\smallskip}
Catalog & The arithmetic mean of $\tau/\rho$ & The value of the distance $\rho$, AU\\
\noalign{\smallskip}\hline\noalign{\smallskip}
$\mathrm{\Phi}$    & $0.108883$ & $\rho<+\infty$        \\
$\mathrm{\Phi}_1$  & $0.100351$ & $\rho\leqslant0.01$   \\
$\mathrm{\Phi}_2$  & $0.100180$ & $\rho\leqslant0.005$  \\
$\mathrm{\Phi}_3$  & $0.100085$ & $\rho\leqslant0.0026$ \\
\noalign{\smallskip}\hline
\end{tabular}
\end{table}

According to our general observations, in the Main Belt an inequality $\tau/\rho>1/2$
is rather rarely satisfied. We observe big values ($0.4$ -- $0.5$) of $\tau/\rho$ mainly
in pairs having a significant mutual inclination and relatively large eccentricities.
Usually those pairs are composed of
asteroids belonging to, for instance, Centaurs or Hungaria family. In~Table~\ref{tab:mean}
we give an averaged value (the usual arithmetic mean) of the ratio $\tau/\rho$ for the
original sample $\mathrm{\Phi}$ and for its subcatalogs $\mathrm{\Phi}_1$, $\mathrm{\Phi}_2$,
$\mathrm{\Phi}_3$. An analysis of~Table~\ref{tab:mean} leads to curious observation:
the mean value of $\tau/\rho$ is very close to $1/10$ and becomes even closer to $1/10$
as the value of $\delta$ decreases.

\begin{table}
\caption{Averaged values of the ratios $\tau/\rho$ and $\sigma/\rho$ for catalogs $\mathrm{\Psi}_k$}
\label{tab:incl}
\begin{tabular}{llll}
\hline\noalign{\smallskip}
$\mathrm{\Psi}_k$ &
\begin{tabular}{l}
The arithmetic\\
mean of $\tau/\rho$
\end{tabular} &
\begin{tabular}{l}
The arithmetic\\
mean of $\sigma/\rho$
\end{tabular} &
\begin{tabular}{l}
The relative position\\
of the orbital planes
\end{tabular} \\[3pt]
\noalign{\smallskip}\hline\noalign{\smallskip}
$\mathrm{\Psi}_1$ & \;\;\;$0.080184$ & \;\;\;$2.358073$ & \;\;\;$0<\sin I\leqslant\sin\frac{\pi}{18}$ \\[3pt]
$\mathrm{\Psi}_2$ & \;\;\;$0.141276$ & \;\;\;$1.184089$ & \;\;\;$\sin\frac{\pi}{18}<\sin I\leqslant\frac12$ \\[3pt]
$\mathrm{\Psi}_3$ & \;\;\;$0.260620$ & \;\;\;$1.075679$ & \;\;\;$\sin I>\frac12$ \\
\noalign{\smallskip}\hline
\end{tabular}
\end{table}

The notion of the distance between two skew lines is the foundation of all constructions
made in the work, so that the appearance of $\sin I$ in the numerator of~\eqref{5-c} is
natural (see the numarator of~\eqref{3-k}). It follows that the lower bound~$\tau$ is
heavily dependent on the mutual inclination, and that the efficiency of the estimate~$\rho\geqslant\tau$
decreases as the orbital configuration approaches the coplanar one. To illustrate this point,
we constructed three different samples $\mathrm{\Psi}_1$, $\mathrm{\Psi}_2$, $\mathrm{\Psi}_3$,
containing one million of pairs each. We put in the catalog $\mathrm{\Psi}_1$ only those
pair configurations, where the angle $I'\stackrel{\mathrm{def}}{=}\min\{I, \pi-I\}$ between
nonoriented orbital planes does not exceed $10^\circ$.
The sample $\mathrm{\Psi}_2$ contains only those pairs that have $10^\circ<I'\leqslant30^\circ$.
In the catalog $\mathrm{\Psi}_3$ each pair has $I'>30^\circ$. The samples
$\mathrm{\Psi}_1$, $\mathrm{\Psi}_2$, $\mathrm{\Psi}_3$ have been composed of arbitrary pairs of
real orbits (we used for constructing $\mathrm{\Psi}_1$, $\mathrm{\Psi}_2$, $\mathrm{\Psi}_3$ all
numbered objects of MPC database) and the value of $I'$ was the only criterion for
compiling of these catalogs. This time along with an averaged value of $\tau/\rho$ we also
calculated the arithmetic mean of the ratio $\sigma/\rho$. Simple geometric considerations
(see~Fig.~\ref{fig:1} and definitions~\eqref{2-sgm},~\eqref{5-tau} of $\sigma$ and $\tau$
respectively) lead to intuitive inference regarding the behaviour of $\tau/\rho$ and $\sigma/\rho$:
the more $I'$, the closer to unity the (mean) ratios $\tau/\rho$ and $\sigma/\rho$ should
be\footnote{Indeed, if the planes of confocal ellipses are far from coplanar configuration,
then the line segment representing the MOID most likely (especially when $e$, $e'$ are small)
is located near the mutual line of nodes. It follows that $\rho$ and $\sigma$ are expected to
approach each other ($\sigma/\rho$ tends to unity) as the orbital configuration approaches
the perpendicular one. Further, the function $C(e, e', I)$ in~\eqref{5-finest} obviously
increases when $I\to\pi/2$ (provided $e$ and $e'$ are hold fixed), which means that $\tau$ is also
expected to increase as $I\to\pi/2$.}. These assumptions are confirmed by Table~\ref{tab:incl},
where we present our mean values of $\tau/\rho$ and $\sigma/\rho$ for each $\mathrm{\Psi}_k$.

Unlike $\tau$, other two bounds $\sigma$ and $\zeta$ considered in this work
(see definitions~\eqref{2-sgm} and~\eqref{1-est} respectively) are certainly optimal.
Indeed, for circular perpendicular orbits ($e=e'=\cos I=0$) with radii~$a$ and~$a'$
we obviously have
$$
\rho=\sigma=\zeta=|a-a'|.
$$
Another simple but less obvious example of configuration with $\rho=\sigma$ is given
by two linked orbits with $a=a', I=\pi/2, \mathbf{P\cdot P'}=-1, e=e'=\varepsilon$,
where $\varepsilon$ is a some small positive number. It is easy to check that if
$\varepsilon$ is sufficiently small, then $\rho=\sigma=2a\varepsilon$.
Note that for linked orbits one always has $\zeta<0$, though the converse statement is not true.

\begin{table}
\caption{The values of $\tau, \zeta, \rho$ and $I$ for some planet pairs in the Solar System}
\label{tab:SS}
\begin{tabular}{lllll}
\hline\noalign{\smallskip}
Pair of planets $\mathcal{E}$ and $\mathcal{E}'$ & $\tau$, AU
                                                 & $\zeta$, AU
                                                 & $\rho$, AU
                                                 & $I$ \\
\noalign{\smallskip}\hline\noalign{\smallskip}
Earth\;  --\; Mars    & $0.007339$ & $0.364732$ & $0.372756$ & $1.8496^\circ$ \\
Uranus\; --\; Neptune & $0.127079$ & $9.806972$ & $9.844646$ & $1.5081^\circ$ \\
\noalign{\smallskip}\hline
\end{tabular}
\end{table}

In conclusion let us notice that $\zeta$ does not depend on the mutual inclination~$I$.
It implies that in some evident cases~---~particularly when $e, e', I$ are quite small~---~the
lower bound~$\zeta$ can be much tighter than~$\tau$. First of all we mean here configurations
composed of main planets of our Solar System, whose orbits have significantly different physical
sizes. Taking the values of the elements from \citep{EMP} we compare in~Table~\ref{tab:SS}
lower bounds $\tau$ and $\zeta$ of the distance $\rho$ for two such configurations.

\section*{Appendix: The distance between two V-sets}
The proof of the following lemma is very simple and therefore is omitted.

\medskip

\emph{Lemma} Let $\mathcal{S}_1, \mathcal{S}_2\subset\mathbb{R}^3$ be two
arbitrary sets satisfying the following three conditions:

i) $\rho(\mathcal{S}_1, \mathcal{S}_2)>0$.

ii) There are two points $Q_1\in\mathcal{S}_1$
    and $Q_2\in\mathcal{S}_2$ such that $\rho(\mathcal{S}_1, \mathcal{S}_2)=Q_1Q_2$.

iii) The pair $(Q_1, Q_2)$ is the only element
     of a set $\mathcal{S}_1\times\mathcal{S}_2$ that
     gives $\rho(\mathcal{S}_1, \mathcal{S}_2)=Q_1Q_2$.
        
Then, for any two subsets $\mathcal{S'}_1\subset\mathcal{S}_1$ and
$\mathcal{S'}_2\subset\mathcal{S}_2$ such that $\mathcal{S'}_1\ni Q_1$ and
$\mathcal{S'}_2\ni Q_2$ one always
has~$\rho(\mathcal{S'}_1, \mathcal{S'}_2)=\rho(\mathcal{S}_1, \mathcal{S}_2)$.

\medskip

For example, if two points $M$ and $N$ lie on skew lines $m$ and $n$
respectively and satisfy $\rho(m, n)=MN$, then for any two rays $m'$
and $n'$ (open or closed, no matter) such that $m'\subset m$, $n'\subset n$,
$m'\ni M$, $n'\ni N$ we have $\rho(m', n')=\rho(m, n)$.

\begin{figure}
\includegraphics[width=0.75\textwidth]{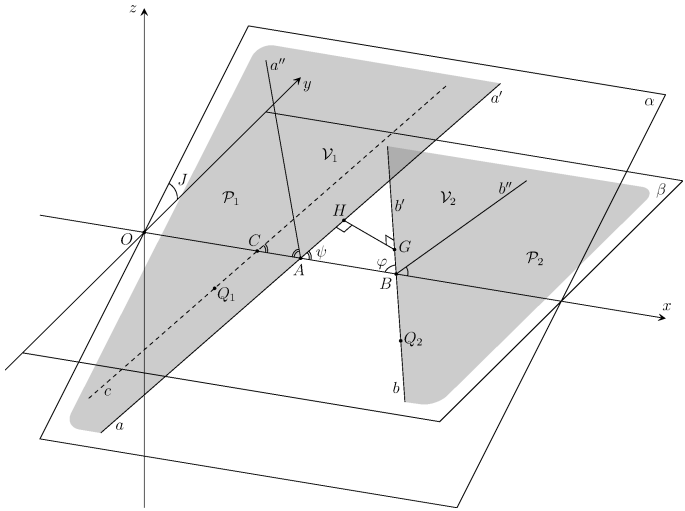}
\caption{The distance between half-planes $\mathcal{P}_1$ and $\mathcal{P}_2$
         is equal to the distance between their edges $a$ and $b$.
         This fact results from elementary similarity and
         continuity considerations ($AB\to0$ implies
         $\rho(\mathcal{P}_1, \mathcal{P}_2)\to0$).
         See text for the strict proof}
\label{fig:10}
\end{figure}

Consider again two two-dimensional closed half-planes $\alpha\subset\{z\geqslant0\}$
and $\beta=\{y\geqslant0; z=0\}$ that form a dihedral angle in $\mathbb{R}^3$ with
the plane angle~$J$ satisfying $0<J\leqslant\pi/2$ (see~Fig.~\ref{fig:10}). In the positive
side of the axis $Ox$ draw points $A$ and $B$ such that $OA<OB$. Given any
positive acute angles~$\psi$ and~$\varphi$, define in the face~$\alpha$ a V-set
$\mathcal{V}_1$ with a vertex $A$ and an exterior angle $\psi$, and in the
face $\beta$ construct a~V-set $\mathcal{V}_2$ with a vertex $B$ and an
exterior angle $\varphi$ (see~Fig.~\ref{fig:10}). Decompose boundaries of~$\mathcal{V}_1$
and~$\mathcal{V}_2$ into four closed rays $a', a'', b', b''$, where
$a', a''\subset\mathcal{V}_1$, $b', b''\subset\mathcal{V}_2$, in such a way that $a'', b'$
both intersect the plane $\{x=0\}$. Further, draw
two straight lines $a$, $b$ such that
$a\supset a'$ and $b\supset b'$. The line $a$ defines
in the plane of $\mathcal{V}_1$ a closed half-plane $\mathcal{P}_1$ that (completely)
contains~$\mathcal{V}_1$. Similarly, define a closed half-plane~$\mathcal{P}_2$
with the edge $b$ such that $\mathcal{P}_2\supset\mathcal{V}_2$ (see~Fig.~\ref{fig:10}).
Our aim is to prove that
\begin{equation}\label{a-vv}
\rho(\mathcal{V}_1, \mathcal{V}_2)=\rho(a, b).
\end{equation}

First of all, draw two points $H\in a$ and $G\in b$ that give $\rho(a, b)=HG$.
It is easy to check, that under the conditions
$$
0<\varphi,\psi<\pi/2,\qquad0<J\leqslant\pi/2
$$
we always have $H_z, G_y>0$. This yields $H\in a'$, $G\in b'$ and hence
we can write
\begin{equation}\label{a-abvv}
\mathcal{V}_1\ni H, \mathcal{V}_2\ni G, a\ni H, b\ni G.
\end{equation}

Further, show that $\mathcal{P}_1$ and $\mathcal{P}_2$ satisfy all conditions
of Lemma. For this, we prove that
\begin{equation}\label{a-rpp}
\rho(\mathcal{P}_1, \mathcal{P}_2)=HG
\end{equation}
and verify that a pair $(H, G)$ is the only element of a set
$\mathcal{P}_1\times\mathcal{P}_2$ satisfying~\eqref{a-rpp}.
Fix any pair $(Q_1, Q_2)\in\mathcal{P}_1\times\mathcal{P}_2$
distinct from the pair $(H, G)\in\mathcal{P}_1\times\mathcal{P}_2$.
It suffices to prove that~$Q_1Q_2>HG$. There are three possibilities.

A) $Q_1\in a$, $Q_2\in b$.

B) One of the points $Q_1, Q_2$ is interior for the half-plain containing it,
   while another is boundary~one.

C) $Q_1\notin a$, $Q_2\notin b$.

Consider case A). Since straight lines $a$ and $b$ are skew, one concludes $Q_1Q_2>HG$.

Consider case B). Let, for example, $Q_1\notin a$, $Q_2\in b$~(see~Fig.~\ref{fig:10}). Through
the point $Q_1$ draw a straight line $c\parallel a$ and denote by $C$ a point where
$c$ (dashed in Fig.~\ref{fig:10}) meets the axis $Ox$. We have
$$
Q_1Q_2\geqslant\rho(c,b)=K(\psi, \varphi, J)CB>K(\psi, \varphi, J)AB=\rho(a,b)=HG,
$$
and therefore $Q_1Q_2>HG$.

Case C) differs from case B) only in that we have to draw auxiliary straight lines
in both half-planes~$\mathcal{P}_1$ and~$\mathcal{P}_2$.

We see that $\mathcal{P}_1$ and $\mathcal{P}_2$ satisfy all conditions of Lemma.
In view of~\eqref{a-abvv}
by~Lemma we conclude that
$$  
  \rho(\mathcal{V}_1, \mathcal{V}_2)=\rho(\mathcal{P}_1, \mathcal{P}_2),\qquad
  \rho(a, b)=\rho(\mathcal{P}_1, \mathcal{P}_2),
$$
which finally implies~\eqref{a-vv}.

\begin{acknowledgements}
We are grateful to Professor K. V. Kholshevnikov for the statement of the problem,
for important remarks and for his help in preparing the manuscript. We also express
gratitude to A. Ravsky for valuable discussion as well as unknown reviewers, whose
constructive and valuable comments greatly helped the authors to improve the manuscript.
All calculations made in the work
were conducted by means of the equipment of the Computing Centre of Research Park
of Saint Petersburg State University. This work is supported by the Russian Science
Foundation grant no. 18-12-00050.
\end{acknowledgements}

\section*{Compliance with Ethical Standards}
{\bf Conflict of interest:} The authors declare that they have no conflicts of interest.
{\bf Ethical approval:} This article does not contain any studies with human
participants or animals performed by any of the authors.
{\bf Informed consent:} This research did not involve human participants.

\bibliographystyle{spbasic}       
\bibliography{mikrbal}            

\end{document}